\documentclass[12pt,a4paper,epsf]{article}
\usepackage{graphics}
\usepackage{amssymb,amsmath}
\usepackage[dvips]{lscape,graphicx}
\usepackage{cite}

\newcommand{\ct}{\cite}
\newcommand{\lb}{\label}

\newcommand{\bc}{\begin{center}}
\newcommand{\ec}{\end{center}}
\newcommand{\bd}{\begin{displaymath}}
\newcommand{\ed}{\end{displaymath}}
\newcommand{\be}{\begin{equation}}
\newcommand{\ee}{\end{equation}}
\newcommand{\ba}{\begin{array}}
\newcommand{\ea}{\end{array}}
\newcommand{\bea}{\begin{eqnarray}}
\newcommand{\eea}{\end{eqnarray}}
\newcommand{\bt}{\begin{tabular}}
\newcommand{\et}{\end{tabular}}

\newcommand{\ov}{\overline}
\newcommand{\bp}{\begin{picture}}
\newcommand{\ep}{\end{picture}}
\newcommand{\bfi}{\begin{figure}}
\newcommand{\efi}{\end{figure}}

\begin{document}

\hyphenation{ }

\title{\huge\bf Composite model of quark-leptons and duality}

\author{\large C.R.~Das ${}^{1}$ \footnote{\large\, crdas@imsc.res.in}\,\,
and Larisa Laperashvili ${}^{1,\, 2}$ \footnote{\large\, laper@itep.ru, 
laper@imsc.res.in}\\[5mm]
\itshape{${}^{1}$ The Institute of Mathematical Sciences, Chennai, India}
\\[0mm]
\itshape{${}^{2}$ The Institute of Theoretical and Experimental Physics, Moscow,
 Russia}}

\date{}

\maketitle

\vspace{1cm}

\bc
{\Large\bf Prepared for the\\[5mm]
17th DAE-BRNS High Energy Physics Symposium,\\[5mm]
IIT, Kharagpur, India, 11-15 December, 2006}\\[10mm]
{\huge \bf Speaker: Chitta Ranjan Das}
\ec

\thispagestyle{empty}

\clearpage\newpage

\thispagestyle{empty}

\begin{abstract}

In the present investigation the model of preons and their composites is 
constructed in the framework of the superstring-inspired flipped
$E_6\times \widetilde{E_6}$ gauge group of symmetry which reveals 
a generalized dual symmetry.
Here $E_6$ and $\widetilde{E_6}$ are non-dual and dual sectors of theory 
with hyper-electric $g$ and hyper-magnetic $\tilde g$ charges, respectively.
Considering preons belonging to the 27-plet of $E_6$, we follow the old idea 
by J.~Pati: we assume that preons are dyons, which in our model are confined
by hyper-magnetic strings -- composite ${\rm\bf N}=1$ supersymmetric non-Abelian flux 
tubes created by the condensation of spreons near the Planck scale. 
Investigating the breakdown of $E_6$ and $\widetilde{E_6}$ at the Planck scale 
into the $SU(6)\times U(1)$ gauge group, we show that the six types of strings 
having fluxes $\Phi_n=n\Phi_0$ $(n=\pm 1,\pm 2,\pm 3)$ produce three 
(and only three) generations of composite quark-leptons and bosons. 
Assuming the existence of the three types of hyper-magnetic fluxes and 
using Schwinger's formula, we have given an explanation of hierarchies of 
masses established in the Standard Model (SM). The following values of masses 
obtained in our preonic model: 
$$ m_t\approx 173\,\,{\mbox{GeV}},\quad m_c\approx 1 \,\,{\mbox{GeV}}
         \quad {\mbox{and}}\quad m_u\approx 4 \,\,{\mbox{MeV}}, 
$$
$$ 
  m_b \approx 4\,\,{\mbox{GeV}},\quad m_s\approx 140 \,\,{\mbox{MeV}}
         \quad {\mbox{and}}\quad m_d\approx 4 \,\,{\mbox{MeV}}, 
$$
$$ 
 m_{\tau}\approx 2\,\,{\mbox{GeV}}\quad {\mbox{and}}\quad m_{\mu}
\approx 100 \,\,{\mbox{MeV}},
$$
are in agreement with the experimentally known results.

The following left-handed neutrino masses were predicted:
$$
     m_1\approx 1.3\times 10^{-3}\,\,{\rm eV}, \quad
 m_2\approx 9.2\times 10^{-3}\,\,{\rm eV}, \quad
 m_3\approx 5.0\times 10^{-2}\,\,{\rm eV} \quad 
$$
-- for direct hierarchy,
$$
     m_1\approx 0.73\times 10^{-2}\,\,{\rm eV}, \quad
 m_2\approx 7.4\times 10^{-2}\,\,{\rm eV}, \quad
 m_3\approx 5.5\times 10^{-2}\,\,{\rm eV} \quad $$ -- for inverted hierarchy.

The compactification in the space-time with five dimensions and its influence 
on form-factors of composite objects are briefly discussed.

\end{abstract}

\pagenumbering{arabic}

\clearpage\newpage

\thispagestyle{empty}

\bc
     {\huge \bf Contents:}
\ec

{\Large \bf

\begin{itemize}

\item[1. ] Introduction: Superstring $E_8\times E'_8$ theory.\\

\item[2. ] Supersymmetric `flipped' $E_6$-unification of gauge interactions.\\

\item[3. ] A new preonic model :\\ 
Preons are dyons confined by hyper-magnetic strings.\\

\item[4. ] Naturalness of three generations.\\

\item[5. ] 
An explanation of mass hierarchies in the Standard Model.\\

\item[6. ] The prediction of neutrino masses.\\

\item[7. ] Form factors as an indication  of the compositeness of 
quark-leptons.

\end{itemize}}

\clearpage\newpage

\setcounter{page}{1}

\begin{itemize}

\item[{\bf 1.}] An explicit construction of model of preons and their composites, 
unified in the supersymmetric flipped $E_6$ GUT, is prescribed in the 
framework of superstring-inspired flipped $E_6\times \widetilde{E_6}$
gauge group of symmetry, which reveals a generalized dual symmetry.
Here $E_6$ and $\widetilde{E_6}$ are non-dual and dual sectors of theory 
with hyper-electric $g$ and hyper-magnetic $\tilde g$ charges, respectively.

Pati was first \ct{1} who suggested to use the strong $U(1)$ magnetic force
as the binding force for preons-dyons making the composite objects.
This idea was developed in Refs.~\ct{2} and has an extension in our model 
in the light of recent investigations of composite non-Abelian flux tubes 
in SQCD \ct{3,4,4a,4b}.

We assume that preons are dyons, which in our model \ct{5} are confined
by hyper-magnetic strings -- composite ${\rm\bf N}=1$ supersymmetric non-Abelian flux 
tubes created by the condensation of spreons near the Planck scale. 

\item[{\bf 2.}] Considering the role of compactification we start, as in 
Ref.~\ct{6}, with ${\rm\bf N}=1$ supersymmetric gauge theory in $5D$ dimensions 
with a local symmetry gauge group $G$ which in our case is equal to the 
flipped $E_6$. The matter fields transform according to one of irreducible 
representation of $G$. For example, we have a 27-plet of preons similar to 
the fundamental 27 representation for quarks and leptons, which decompose under 
$SU(5)\times U(1)$ subgroup as follows: 
\be
        27 \to (10,1) + \left(\bar 5, -3\right) +  \left(\bar 5,2\right) +
          (5,-2) + (1,5) + (1,0).                             \lb{3}
\ee
The first and second quantities in the brackets of Eq.~(\ref{3}) correspond to 
the $SU(5)$ representation and $U(1)_X$ charge, respectively. 

The conventional SM family which contains the doublets of left-handed quarks 
$Q$ and leptons $L$, right-handed up and down quarks $u^{\rm\bf c}$, $d^{\rm\bf c}$, 
also $e^{\rm\bf c}$,  
is assigned to the $(10,1) + (\bar 5,-3) + (1,5)$ representations 
of the flipped $SU(5)\times U(1)_X$, along with right-handed neutrino $N^{\rm\bf c}$. 
These representations decompose under
\be
SU(5)\times U(1)_X \to SU(3)_C\times SU(2)_L\times U(1)_Z\times U(1)_X,      
                                                     \lb{4a}
\ee
We consider charges $Q_X$ and $Q_Z$ in the units ${1}/{\sqrt{40}}$ 
and $\sqrt{3/5}$, respectively, using assignments: $Q_X=X$ and $Q_Z=Z$ 
(see \ct{11} for details).

The decomposition (\ref{4a}) gives the following content:

\bea
       (10,1) \to Q = &\left(\begin{array}{c}u\\ 
                                          d \end{array}\right) &\sim 
                         \left(3,2,\frac 16,1\right),\nonumber\\
&d^{\rm\bf c} &\sim \left(\bar3,1,-\frac 23,1\right),\nonumber\\
&N^{\rm\bf c} &\sim \left(1,1,1,1\right).       \lb{4}\\
\left(\bar 5,-3\right) \to &u^{\rm\bf c}&\sim \left(\bar 3,1,\frac 13,-3\right),\nonumber\\
L = &\left(\begin{array}{c}e\\ 
                                             \nu \end{array}\right) &\sim 
                         \left(1,2,-\frac 12,-3\right),               \lb{5}\\
(1,5) \to &e^{\rm\bf c} &\sim \left(1,1,1,5\right).\lb{6}
\eea

The remaining representations in Eq.~(\ref{3}) decompose as follows:

\bea
        (5,-2) \to& D&\sim \left(3,1,-\frac 13,-2\right),\nonumber\\
                   h = &\left(\begin{array}{c}h^+\\ 
                                               h^0 \end{array}\right) &\sim 
                         \left(1,2,\frac 12,-2\right).
                                                              \lb{7}\\
    \left(\bar 5,2\right) \to &D^{\rm\bf c} &\sim \left(\bar 3,1,\frac 13,2\right),\nonumber\\
                     h^{\rm\bf c} = &\left(\begin{array}{c}h^0\\ 
                                               h^- \end{array}\right) &\sim 
                         \left(1,2,-\frac 12,2\right).              \lb{8}
\eea
The light Higgs doublets are accompanied by coloured Higgs triplets $D,D^{\rm\bf c}$.

The singlet field $S$ is represented by (1,0):
\be
       (1,0) \to S \sim (1,1,2,2).               \lb{9}
\ee
It is necessary to notice that the flipping of our $SU(5)$:
\be
       d^{\rm\bf c} \leftrightarrow u^{\rm\bf c},\quad 
N^{\rm\bf c}\leftrightarrow e^{\rm\bf c}, \lb{10}
\ee
distinguishes this group of symmetry from the standard Georgi-Glashow $SU(5)$.

Supermultiplet ${\cal V} = (A^M, \lambda^{\alpha}, \Sigma, X^a)$ in 
$5D$-dimensional space (with $M=0,1,2,3,4$ space-time indices) contains 
a vector field:
\be
            A^M = A^{MJ}T^J,     \lb{13}                                      
\ee
and a real scalar field:
\be
                  \Sigma = \Sigma^JT^J,             \lb{14}
\ee
where $J$ runs over the $E_6$ group index values and $T^J$ are generators 
of $E_6$ algebra.

Two gauginos fields:
\be
            \lambda^{\alpha} = \lambda^{\alpha J}T^J,        \lb{15}
\ee
form a decuplet under the $R$-symmetry group $SU(2)_R$ (with $\alpha=1,2$).

Auxiliary fields:
\be     
                  X^a = X^{aJ}T^J                      \lb{16}
\ee
form a triplet of $SU(2)_R$ (with $a=1,2,3$).

After compactification, these fields are combined into the ${\rm\bf N}=1$ $4D$ fields:
$$
   {\mbox{vector supermultiplet}}\quad V=\left(A^{\mu}, \lambda^1, X^3\right)
$$
(here $\mu=0,1,2,3$ are the space-time indices), and
$$
 {\mbox{a chiral supermultiplet}}\quad \Phi=\left(\Sigma+iA^4, \lambda^2, 
X^1+iX^2\right).
$$
The matter fields are contained in the hypermultiplet:
$$
       {\cal H} = (h^{\alpha}, \Psi, F^{\alpha}),
$$
where $h^{\alpha}$ is a doublet of $SU(2)_R$, Dirac fermion field 
$\Psi=(\psi_1, \psi_2^+)^T$
is the $SU(2)_R$ singlet and auxiliary fields $F^{\alpha}$ also are the 
$SU(2)_R$ doublet.

Then we have two ${\rm\bf N}=1$ $4D$ chiral multiplets:
$$
              H = \left(h^1, \psi_1, F^1\right)\quad {\rm and}\quad H^{\rm\bf c} = 
                                 \left(h^2, \psi_2, F^2\right)  
$$ 
transforming according to the representations $R$ and anti-$R$ of the gauge group 
$G=E_6$, respectively.
Then the $5D$-dimensional $E_6$-symmetric action is given as 
follows:
\be
       S = \int d^5x\int d^4\theta\left[H^{\rm\bf c}e^VH^{{\rm\bf c}+} + H^+e^VH\right] + 
        \int d^5x\int d^2\theta\left[H^{\rm\bf c} \left(\partial_4 - 
\frac 1{\sqrt 2}\Phi\right)H 
                                + h.c.\right].   \lb{17}           
\ee
This theory is anomaly-free (see \ct{6} and references therein).

Compactifying the extra fifth dimension $x^4$ in Eq.~(\ref{17}) on 
a circle of radius $R_C$, the authors of Ref.~\ct{6} impose the Scherk-Schwarz 
boundary conditions to the preonic superfields and obtained the 
following result:
\bea
   P\left(x^m, x^4+2\pi R_C, \theta\right) &=& e^{i2\pi q_P}P\left(x^m, x^4, e^{i\pi
(q_P+q_{\bar P})}\theta\right),                  
\nonumber\\
   P^c\left(x^m, x^4+2\pi R_C, \theta\right) &=& e^{i2\pi q_{\bar P}}P^c\left(x^m, x^4, e^{i\pi
(q_P+q_{\bar P})}\theta\right),                  
\nonumber\\
   P_s\left(x^m, x^4+2\pi R_C, \theta\right) &=& e^{i2\pi q_s}P_s\left(x^m, x^4, e^{i\pi
(q_s+q_{\bar s})}\theta\right),                  
\nonumber\\
   P^c_s\left(x^m, x^4+2\pi R_C, \theta\right) &=& e^{i2\pi q_{\bar s}}P^c_s
\left(x^m, x^4, e^{i\pi(q_s+q_{\bar s})}\theta\right),                  \lb{a1}
\eea
where $q_P$, $q_{\bar P}$ and  $q_s$, $q_{\bar s}$ are $R$ charges of preons 
$P$, $P^c$ and $P_s$, $P^c_s$, respectively.
 The following conditions were obtained:
 \be   
      q_P+q_{\bar P} = q_s+q_{\bar s}.              \lb{a2}
\ee
As a result of compactification, all fermionic preons are massive in $4D$ 
space-time. Supersymmetry is broken by the boundary conditions (\ref{a1}).  

\item[{\bf 3.}]
Why three generations exist in Nature ? We suggest an 
explanation considering a new preonic model of composite SM particles.
The model starts from the supersymmetric flipped $E_6\times \widetilde {E_6}$ 
gauge group of symmetry for preons, where  $E_6$ and $\widetilde {E_6}$ 
correspond to non-dual (with hyper-electric charge $g$) and dual (with 
hyper-magnetic charge $\tilde g$) sectors of theory, respectively. 
We assume that preons are dyons confined by hyper-magnetic strings in the 
region of energy $\mu < M_{Pl}$.

{\Large\bf Preons are dyons bound by hyper-magnetic strings}

Considering the ${\rm\bf N}=1$ supersymmetric flipped $E_6\times \widetilde {E_6}$
gauge theory for preons in $4D$-dimensional space-time, we assume that preons 
$P$ and antipreons $P^{\rm\bf c}$ are 
dyons with charges $g$ and $\tilde g$, respectively, resided
in the $4D$ hypermultiplets ${\cal P} = (P,P^{\rm\bf c})$ and ${\cal \tilde P} = 
(\tilde P, {\tilde P}^{\rm\bf c})$. Here ``$\tilde P$" designates spreons, but not 
the belonging to $\widetilde{E_6}$.

The dual sector $\widetilde{E_6}$ is broken in our world to some group $\widetilde{G}$,
and preons and spreons transform under the hyper-electric gauge group $E_6$ 
and hyper-magnetic gauge group $\widetilde{G}$ as their fundamental representations:
\be
   P,\tilde P\sim (27,N), \quad  P^{\rm\bf c},\tilde P^{\rm\bf c}\sim 
\left(\ov {27},\bar N\right),
                                         \lb{19}
\ee
where $N$ is the $N$-plet of $\widetilde{G}$ group.
We also consider scalar preons and spreons as singlets of $E_6$:
\be
   P_s,\tilde P_s\sim (1,N), \quad  P_s^{\rm\bf c},\tilde 
P_s^{\rm\bf c}\sim \left(1,\bar N\right),
                                         \lb{20}
 \ee  
which are actually necessary for the entire set of composite 
quark-leptons and bosons. 

The hyper-magnetic interaction is assumed to be responsible for the formation
of $E_6$ fermions and bosons at the compositeness scale $\Lambda_s$.
The main idea of the present investigation is an assumption that
preons-dyons are confined by hyper-magnetic supersymmetric 
non-Abelian flux tubes which are a generalization of the well-known 
Abelian ANO-strings for the case of the supersymmetric 
non-Abelian theory developed in Refs.~\ct{3,4,4a,4b}.
As a result, in the limit of infinitely narrow flux tubes (strings) we have 
the following bound states:

\begin{itemize}
\item[{\bf i.}] quark-leptons (fermions belonging to the $E_6$ fundamental representation):
\bea
            Q^a &\sim& P^{aA}(y) \left[{\cal P}\exp\left( i\tilde g 
\int_x^y\widetilde{A}_{\mu}dx^{\mu}\right)\right]_A^B
                         (P_s^{\rm\bf c})_B(x) \sim 27,
                                               \lb{21}
\\
          \bar  Q_a &\sim& (P_s^{\rm\bf c})^A(y) 
\left[{\cal P}\exp\left( i\tilde g \int_x^y\widetilde{A}_{\mu}dx^{\mu}\right)\right]_A^B
                         P_{aB}(x) \sim \ov {27},
                                               \lb{22}
\eea     
where $a\in 27$-plet of $E_6$, $A, B\in N$-plet of $\widetilde{G}$, and 
$\widetilde{A}_{\mu}(x)$ are dual hyper-gluons belonging to the adjoint 
representation of $\widetilde{G}$;

\item[{\bf ii.}] ``mesons" (hyper-gluons and hyper-Higgses of $E_6$):
\bea
            M^a_b &\sim& P^{aA}(y) \left[{\cal P}\exp\left( i\tilde g 
\int_x^y\widetilde{A}_{\mu}dx^{\mu}\right)\right]_A^B
                         (P^{\rm\bf c})_{bB}(x)\nonumber\\ 
&&\sim 1+78+650\quad {\rm of}\,\,E_6,
                                               \lb{23}
\\
            S &\sim& (P_s)^A(y) \left[{\cal P}\exp\left( i\tilde g 
\int_x^y\widetilde{A}_{\mu}dx^{\mu}\right)\right]_A^B
                         (P_{s}^{\rm\bf c})_B(x) \sim 1,
                                               \lb{24}
\\
          \bar S &\sim& (P_s^{\rm\bf c})^A(y) 
\left[{\cal P}\exp\left( i\tilde g 
\int_x^y\widetilde{A}_{\mu}dx^{\mu}\right)\right]_A^B
                         (P_s)_B(x) \sim 1;
                                               \lb{25}
\eea   
\clearpage\newpage
\item[{\bf iii.}] ``baryons",

 for $\widetilde{G}$-triplet we have (see Section 5):
\bea
             D_1 &\sim& \epsilon_{ABC}P^{aA'}(z)P^{bB'}(y)P^{cC'}(x)
\left[{\cal P}\exp\left( i\tilde g \int^z_X\widetilde{A}_{\mu}dx^{\mu}\right)\right]_{A'}^A
\times
\nonumber\\ &&
 \left[{\cal P}\exp\left( i\tilde g \int^y_X\widetilde{A}_{\mu}dx^{\mu}\right)\right]_{B'}^B
 \left[{\cal P}\exp\left( i\tilde g \int^x_X\widetilde{A}_{\mu}dx^{\mu}\right)\right]_{C'}^C,      \lb{26}
\\
             D_2 &\sim& \epsilon_{ABC}P^{aA'}(z)P^{bB'}(y)(P_s)^{C'}(x)
\left[{\cal P}\exp\left( i\tilde g \int^z_X\widetilde{A}_{\mu}dx^{\mu}\right)\right]_{A'}^A
\times
\nonumber\\ &&
 \left[{\cal P}\exp\left( i\tilde g \int^y_X\widetilde{A}_{\mu}dx^{\mu}\right)\right]_{B'}^B
 \left[{\cal P}\exp\left( i\tilde g \int^x_X\widetilde{A}_{\mu}dx^{\mu}\right)\right]_{C'}^C,      \lb{27}
\eea 
and their conjugate particles.
\end{itemize}

The bound states (\ref{21})--(\ref{27}) are shown
in Fig.~1 as unclosed strings $(a)$ and ``baryonic'' configurations $(b)$.
It is easy to generalize Eqs.~(\ref{21})--(\ref{27}) for the case of 
string constructions of superpartners -- squark-sleptons, hyper-gluinos 
and hyper-higgsinos. Closed strings -- gravitons -- are presented 
in Fig.~1(c).
All these bound states belong to the $E_6$ representations and they are in 
fact the ${\rm\bf N}=1$ $4D$ superfields.

\item[{\bf 4.}]
In the letter \ct{6a},
it was shown that preonic $E_6$ can be broken by
Higgses belonging to the 78-dimensional representation of $E_6$:
\be
               E_6 \to SU(6)\times SU(2) \to SU(6)\times U(1),
                                                      \lb{28}
\ee
where $SU(6)\times U(1)$ is the largest relevant invariance group of the 78.

$SU(6)\times U(1)$ group of symmetry existing near the Planck scale
is described by theory of non-Abelian tubes in ${\rm\bf N}=1$
SQCD which was developed recently in Refs.~\ct{3,4,4a,4b}.

We assume the condensation of spreons near the Planck scale.
Then theory leads to
the topologically stable string solutions possessing 
both windings, in $SU(6)$ and $U(1)$. Now onwards we assume  
the dual sector of theory described by $\widetilde {SU(6)}\times 
\widetilde {U(1)}$ group of symmetry which produces hyper-magnetic fluxes. 
Then,  according to the results obtained in Refs.~\ct{3,4,4a,4b}, we have 
a nontrivial homotopy group:
\be 
            \pi_1\left(\frac{SU(6)\times U(1)}{Z_6}\right) \neq 0, \lb{29}
\ee
and flux lines form topologically non-trivial $Z_6$ strings.

The model  contains six scalar fields 
charged with respect to $U(1)$ and belonged to the 6-plet of $SU(6)$.
Considering scalar fields of spreons:  
\be
      \tilde P=\left\{ \phi^{aA}\right\},\quad{\mbox {where}}\quad a,A=1,..,6, 
\lb{30}
\ee
we can construct condensation of spreons in vacuum:
\be
       {\tilde P}_{vac} = \left\langle\tilde P^{aA}\right\rangle = 
v\cdot{\mbox{diag}}(1,1,..,1),\quad
                               {\mbox{a,A=1,..,6}}. \lb{31}
\ee

The value $v$ is the vacuum expectation value (VEV), which in theory of 
Refs.~\ct{3} is given by
\be
              v =\sqrt \xi \gg \Lambda_4,           \lb{32} 
\ee
where $\xi$ is the Fayet-Iliopoulos $D$-term parameter in the ${\rm\bf N}=1$ 
supersymmetric theory and $\Lambda_4$ is its 4-dimensional scale.
In our case:
\be
             v\sim M_{Pl}\sim 10^{19}\,\,{\mbox{GeV}},   \lb{33}    
\ee
because spreons are condensed at the Planck scale.

Non-trivial topology (\ref{29}) amounts to winding of elements of matrix 
(\ref{30}), and we obtain string solutions:
\be
 {\tilde P}_{string} = v\cdot{\mbox{diag}}\left(e^{i\alpha(x)},e^{i\alpha(x)},..,1,1\right)
                          \quad {\mbox{where}}\quad x\to \infty. \lb{34}
\ee
Investigating string moduli space (see Ref.~\ct{4,4b}),
we obtain the solutions for six types of $Z_6$-flux tubes which are 
a non-Abelian analog of Abrikosov-Nielsen-Olesen (ANO)-strings.

Considering  at the ends of strings preons $P$ and $P^{\rm\bf c}$ with hyper-magnetic 
charges $n\tilde g$ and $-n\tilde g$, respectively, we obtain six 
types of strings having the fluxes $\Phi_n$ quantized according 
to the $Z_6$ center group of $SU(6)$:
\be
      \Phi_n = n\Phi_0, \quad n=\pm 1,\pm 2,\pm 3.
                                                       \lb{36}          
\ee
String tensions of these non-Abelian flux tubes also were calculated in
Refs.~\ct{3}. The minimal tension is:
\be
               T_0 = 2\pi \xi,                  \lb{38}                  
\ee
which in our preonic model is equal to:
\be
               T_0 = 2\pi v^2\sim 10^{38}\,\, {\rm GeV}^2,
                                                          \lb{39}
\ee
that is, enormously large.
This means that preonic strings have almost infinitely small 
$\alpha' \to 0$, where $\alpha'= 1/(2\pi T_0)$ is a slope of trajectories 
in string theory \ct{1}. Six types of preonic tubes give three types of 
$k$-strings having the following tensions:
\be            
                    T_k =kT_0, \quad {\mbox{where}}\quad k=1,2,3. \lb{39a}
\ee
Thus, hyper-magnetic charges of preons and antipreons are confined by six 
flux tubes which are oriented  
in opposite directions, but have only three different tensions (\ref{39a}). 

Also preonic strings are enormously thin. 
As it was shown in the letter \ct{6a}, the thickness of preonic strings 
given by the radius $R_{str}$ of the flux tubes is very small:
\be
      R_{str}\sim \frac 1{m_V} \sim \frac 1{gv}\sim 10^{-18}
                                          \,\,{\rm GeV}^{-1}. \lb{42}
\ee
Such infinitely narrow non-Abelian supersymmetric flux tubes remind us
superstrings of Superstring theory.

\item[{\bf 5.}] In the previous item we have given a demonstration of a very 
specific type of the ``horizontal symmetry": three, and only three, 
generations of fermions and bosons exist in the superstring-inspired flipped
$E_6$ theory. All bound states (\ref{4})--(\ref{9}) form three generations -- 
three 27-plets of $E_6$.
We also obtain the three types of gauge bosons $A_{\mu}^i$ ($i=1,2,3$
is the index of generations) 
Fig.~2 illustrates the formation of such 
hyper-gluons (Fig.~2(a)) and also hyper-Higgses of Fig.~2(b).

Having in our preonic model  supersymmetric strings with $\alpha'\to 0$ we
obtain, according to the description \ct{1}, only massless ground states:
spin 1/2 fermions (quarks and leptons), spin 1 hyper-gluons and spin 2 
massless graviton, as well as their superpartners. The exited states 
belonging to these strings are not realized in our world as very massive: 
they have mass $M > M_{Pl}$. 

Using our preonic model, we give a simple explanation
why quarks and leptons of three SM generations have such 
different masses. We show that the hierarchy of the SM masses is connected
with the string construction of preon bound states.

New preon-antipreon pairs can be generated in the flux tubes between preons. Our 
assumption is that the mechanism of preon-antipreon pair production in these
tubes is similar to that of $e^+e^-$ pair production in the uniform constant
electric field considered in one space dimension (see \ct{5} and references therein).

Almost thirty years ago J.~Schwinger obtained the following expression for
the rate per unit volume that a $e^+e^-$ pair will be created in the constant 
electric field of strength $\cal E$:
\be
           W = \frac {(e{\cal E})^2}{4\pi^3} \sum_{n=1}^{\infty}\frac 1{n^2}
                     \exp\left(-\frac{\pi m_e^2n}{e{\cal E}}\right),           \lb{52}
\ee
where $e$ is the electron charge and $m_e$ its mass.

A number of authors have applied this result to hadronic productions 
considering a creation of quark-antiquark pairs in the QCD tubes of colour
flux.

Considering that the probability $W_k$ of one preon-antipreon pair
production in a unit space-time volume of the $k$-string tube 
with tension (\ref{39a}) is given by Schwinger's expression:
\be
              W_k = \frac{T_k^2}{4\pi^3}\exp\left(-\frac{\pi M_k^2}{T_k}\right),
            \quad  {\rm where}\quad T_k=kT_0,\quad k=1,2,3,                  \lb{53}
\ee
we assume that Yukawa couplings $Y_k$ for the three generations of quarks 
($t,c$ and $u$) are proportional to the square root of probabilities $W_k$. Then we obtain 
the following ratios:
\be
       Y_t : Y_c : Y_u :: W_1^{1/2} : W_2^{1/2} : W_3^{1/2}.              \lb{54}
\ee
In Eq.~(\ref{53}) the values $M_k$ are the constituent masses of preons 
produced in the $k$-strings. Yukawa couplings $Y_k$ are proportional to 
the masses of quarks $t,c,u$, and we have:
\be
       m_t : m_c : m_u :: \frac{T_0}{2\pi^{3/2}}\exp\left(-\frac{\pi M_1^2}{2T_0}\right) :
           \frac{2T_0}{2\pi^{3/2}}\exp\left(-\frac{\pi M_2^2}{4T_0}\right) :
            \frac{3T_0}{2\pi^{3/2}}\exp\left(-\frac{\pi M_3^2}{6T_0}\right).      \lb{55}
\ee
Assuming that $M_k$ are proportional to $k$:
\be
                  M_k =kM_0,                    \lb{56}
\ee
we get the following result:
\be
 m_t : m_c : m_u = m_0\left(w : 2w^2 : 3w^3\right),      \lb{57}
\ee
where
\be
             w = \exp\left(-\frac{\pi M_0^2}{2T_0}\right),            \lb{58}
\ee
and $m_0$ is a mass parameter.
 
For parameter 
\be
                w=w_1\approx 2.9\cdot 10^{-3}                              \lb{59}
\ee 
we obtain the following values for masses of $t,c,u$-quarks:
\be
         m_t\approx 173\,\,{\rm GeV},\quad m_c\approx 1\,\,{\rm GeV}
         \quad {\rm and}\quad m_u\approx 4\,\,{\rm MeV}.   \lb{60}
\ee                                            
For $b,s,d$-quarks we have different parameters $M_0$, $m_0$ and $w=w_2$.
The value:
\be
                   w_2\approx 1.7\cdot 10^{-2}                \lb{61}
\ee
gives the following masses of $b,s,d$-quarks: 
\be
         m_b\approx 4\,\,{\rm GeV},\quad m_s\approx 140\,\,{\rm MeV}
         \quad {\rm and}\quad m_d\approx 4\,\,{\rm MeV}.   \lb{62}
\ee       
The results (\ref{60}) and (\ref{62}) are in agreement with experimentally 
established quark masses published in Ref.~\ct{7}:
\be
         m_t\approx 174\pm 5.1\,\,{\rm GeV},\quad m_c\approx 1.15  
\,\,{\rm to}\,\,1.35\,\,{\rm GeV}
         \quad {\rm and}\quad m_u\approx 1.5 \,\,{\rm to}\,\,4 
                 \,\,{\rm MeV}.   \lb{63}
\ee          
and
\be
         m_b \approx 4.1\,\,{\rm to}\,\,4.9
    \,\,{\rm GeV},\quad m_s\approx 80\,\,{\rm to}\,\,130
                  \,\,{\rm MeV}
         \quad {\rm and}\quad m_d\approx 4\,\,{\rm to}\,\,8
         \,\,{\rm MeV}.   \lb{64}
\ee      
The value of $w$-parameter:
\be
                   w_3\approx 2.5\cdot 10^{-2}                \lb{65}
\ee
leads to the following values for masses of $\tau,\mu$-leptons and 
electron: 
\be
   m_{\tau}\approx 2\,\,{\rm GeV},\quad m_{\mu}\approx 100\,\,{\rm MeV}
    \quad {\rm and}\quad m_e\approx 3.5 \,\,{\rm MeV}, \lb{66}
\ee              
which are comparable with the results of experiment \ct{7}:  
\be
   m_{\tau}\approx 1.777\,\,{\rm GeV},\quad m_{\mu}\approx 105.66\,\,
{\rm MeV}\quad {\rm and}\quad m_e\approx 0.51\,\,{\rm MeV}. \lb{67}
\ee      
We see that our preonic model explains the hierarchy of masses
existing in the SM.

\item[{\bf 6.}]
The method permits to predict masses of the left-handed neutrinos \ct{8,9}.
Experimental results on solar neutrino and atmospheric neutrino
oscillations from Sudbury Neutrino Observatory (SNO Collaboration) and
the Super-Kamiokande Collaboration have been used to extract the
following parameters \ct{10}:
\bea
\Delta m^2_\odot &=& m_2^2 - m_1^2 \approx 8.3\times 10^{-5}\,\,
                                         {\rm eV}^2,    \lb{n1}
\\
 \Delta m^2_{\rm atm} &=& \left|m_3^2 - m_2^2\right| \approx    2.4\times 10^{-3}\,\,
                                         {\rm eV}^2,    \lb{n2}
\eea
where $m_1$, $m_2$, $m_3$ are the hierarchical left-handed neutrino
effective masses for three families.

The solution of Eqs.~ (\ref{n1}) and (\ref{n2}) leads to the following result:
\be 
                    m_0\approx 0.540\,\,{\rm eV}, \quad w\approx 0.092.    \lb{n3}
\ee
This solution gives the direct hierarchal masses of the left-handed neutrinos: 
\be
   m_1\approx 1.3\times 10^{-3}\,\,{\rm eV}, \quad
 m_2\approx 9.2\times 10^{-3}\,\,{\rm eV}, \quad
 m_3\approx 5.0\times 10^{-2}\,\,{\rm eV}. \lb{n4}
\ee

For the case of inverted hierarchy of neutrino masses we have obtained 
the following prediction:
\be
  m_1\approx 0.73\times 10^{-2}\,\,{\rm eV}, \quad
 m_2\approx 7.4\times 10^{-2}\,\,{\rm eV}, \quad
 m_3\approx 5.5\times 10^{-2}\,\,{\rm eV}.                  \lb{n5}
\ee

\item[{\bf 7.}]
Non point-like behaviour of the fundamental quarks and leptons is related
with the appearance of form factors describing the dependence of cross 
sections in 4-momentum squared $q^2$ for different elementary particle 
processes, for example, in reactions:
$$
    e^+e^- \to e^+e^-,\,\,\mu^+\mu^-,\,\,\tau^+\tau^-,\,\,\gamma\gamma,
$$ or
$$ 
             e^+e^- \to\,\,{\rm hadrons},
$$ 
or in the deep inelastic lepton scattering experiments, etc.

Considering the form factor $F(q^2)$ as a power series in $q^2$, we find:
\be
   F(q^2) = 1 + \frac{q^2}{\Lambda^2} + O\left(\frac 1{\Lambda^4}\right),
               \quad  {\rm where} \quad q^2 \ll \Lambda^2.  
                                                            \lb{68}
\ee
If $\Lambda=\Lambda_4$ (see item 4)
is the energy scale associated with the quark-lepton 
compositeness scale $\Lambda_s$ of the preon interaction, then our 
model predicts:
\be
      \Lambda \sim \Lambda_s\sim 10^{18}\,\,{\rm GeV}. \lb{69}
\ee
Of course, such a scale is not available even for future high energy
colliders.

From a phenomenological point of view it is quite important to understand 
whether the compactification scale lowers down  $\Lambda$ in Eq.~(\ref{68})
to energies accessible for the future accelerators.

Proposing the presence of extra space-time dimensions at small distances
\ct{6} we have a hope that the radius $R_C$ of compactification 
given by Eq.~(\ref{a1}) is larger than the radius $R_s$ of compositeness:
\be
                  R_C  \gg  R_s,      \lb{70}
\ee
and
\be
 \Lambda \sim \Lambda_C \ll 10^{18}\,\,{\rm GeV}. \lb{71}
\ee
According to Ref.~\ct{6} the $R$ charges for the composite states 
(\ref{21})--(\ref{24}) are:
\be Q\sim q_P + q_{\bar s},\quad \bar Q\sim  q_{\bar P} + q_{s},
  \quad M \sim  q_P + q_{\bar P}, \quad S\sim  q_s + q_{\bar s}.    \lb{a3}
\ee
However, this problem needs the more detailed investigations 
in higher dimensional theories.

\end{itemize}

\bc
{\Large \bf Conclusions}\ec

In this talk we have presented:

\begin{itemize}

\item[{\bf i.}] Supersymmetric $E_6$ content in five space-time dimensions and 
the problem of compactification.

\item[{\bf ii.}] Starting with ${\rm\bf N}=1$ supersymmetric $E_6\times \widetilde{E_6}$ 
preonic model of composite quark-leptons and bosons, we have assumed that 
preons are dyons confined by hyper-magnetic strings in the region of energies 
$\mu \lesssim M_{Pl}$. This approach is an extension of the old idea by
J.~Pati \ct{1} who suggested to use the strong magnetic forces to bind 
preons-dyons in composite particles -- quark-leptons and bosons. Our 
model is based on the recent theory of composite non-Abelian flux tubes 
in SQCD which was developed in Refs.~\ct{3,4}.

\item[{\bf iii.}] Considering the breakdown of $E_6$ and $\widetilde{E_6}$ at the 
Planck scale into the $SU(6)\times U(1)$ gauge group we have shown that 
six types of $k$-strings -- composite ${\rm\bf N}=1$ supersymmetric non-Abelian 
flux tubes --
are created by the condensation of spreons near the Planck scale.

\item[{\bf iv.}] We have obtained that six types of strings-tubes have six fluxes
of hyper-magnetic fields quantized according to the $Z_6$ center group of
$SU(6)$:
$$
      \Phi_n = n\Phi_0, \quad n=\pm 1,\pm 2,\pm 3,
$$
and these fluxes produce three (and only three) generations of composite
quark-leptons and bosons.

\item[{\bf v.}] It was shown that in the present model preonic strings are very thin,
with radius
$$
          R_{str}\sim 10^{-18} \,\,{\rm GeV}^{-1},
$$
and their tension is enormously large:
$$  
                   T\sim  10^{38} \,\,{\rm GeV}^2.
$$
They remind superstrings of Superstring theory.
 
\item[{\bf vi.}] By the existence of three types of hyper-magnetic fluxes,
using Schwinger's formula, we have given an explanation of hierarchies of 
masses established in the SM. Our result: 
$$ m_t\approx 173\,\,{\mbox{GeV}},\quad m_c\approx 1 \,\,{\mbox{GeV}}
         \quad {\mbox{and}}\quad m_u\approx 4 \,\,{\mbox{MeV}}, $$
$$
  m_b \approx 4\,\,{\mbox{GeV}},\quad m_s\approx 140 \,\,{\mbox{MeV}}
         \quad {\mbox{and}}\quad m_d\approx 4 \,\,{\mbox{MeV}}, $$
$$
 m_{\tau}\approx 2\,\,{\mbox{GeV}},\quad m_{\mu}\approx 100 \,\,{\mbox{MeV}}
    \quad {\mbox{and}}\quad m_e\approx 3.5 \,\,{\mbox{MeV}}, 
$$
is comparable with experimentally established results (see Ref.~\ct{7}).

\item[{\bf vii.}]

The following left-handed neutrino masses were predicted:
$$
     m_1\approx 1.3\times 10^{-3}\,\,{\rm eV}, \quad
 m_2\approx 9.2\times 10^{-3}\,\,{\rm eV}, \quad
 m_3\approx 5.0\times 10^{-2}\,\,{\rm eV} \quad 
$$
-- for direct hierarchy,
$$
     m_1\approx 0.73\times 10^{-2}\,\,{\rm eV}, \quad
 m_2\approx 7.4\times 10^{-2}\,\,{\rm eV}, \quad
 m_3\approx 5.5\times 10^{-2}\,\,{\rm eV} \quad $$ -- for inverted hierarchy.
These results are in agreement with experiment \ct{7,10}.

\item[{\bf viii.}] We have considered form factors described the compositeness of 
quark-leptons. We have shown that in our preonic model the scale $\Lambda_s$ 
of the quark-lepton compositeness is very large:
$$
\Lambda_s \sim 10^{18}\,\,{\mbox{GeV}},$$
what is not accessible even for future high energy colliders.
Only the compactification procedure of extra space-time dimensions
can help to lower the scale $\Lambda_s$.

\end{itemize}

{\large \bf Acknowledgements:}
C.R.D. thanks a lot Prof.~J.~Pati for interesting discussions during the 
Conference WHEPP-9 (Bhubaneswar, India, January, 2006). 
L.L. sincerely thanks the Institute of Mathematical Sciences (Chennai, India) 
and personally Director of IMSc Prof.~R.~Balasubramanian and  
Prof.~N.D.~Hari Dass for the wonderful hospitality and financial support.
The authors also deeply thank Prof.~J.L.~Chkareuli,
Prof.~F.R.~Klinkhamer and Prof.~G.~Rajasekaran for fruitful discussions and 
advices. 

This work was supported by the Russian Foundation for Basic Research (RFBR), 
project $N^o$ 05--02--17642.

\clearpage\newpage

\bfi
\centering
\includegraphics[height=62mm,keepaspectratio=true]{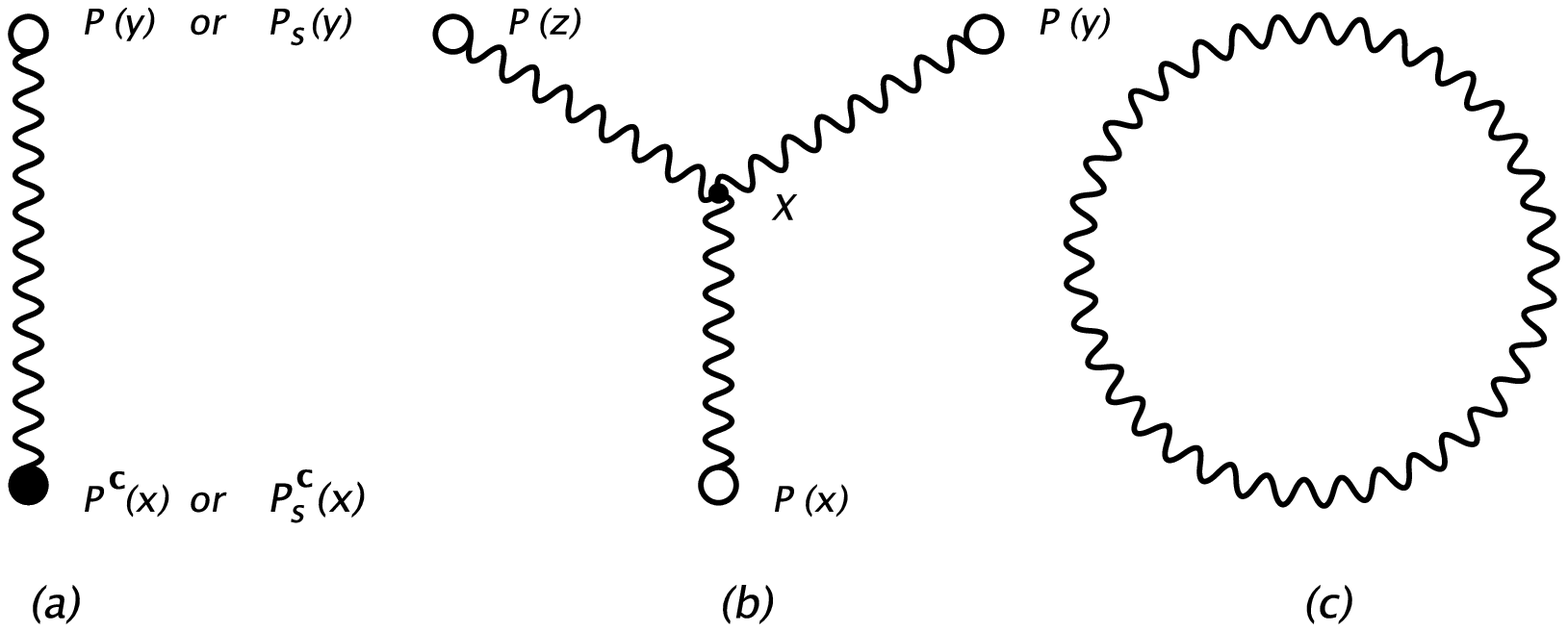}
\caption{Preons are bound by hyper-magnetic strings: (a,b) correspond to
the string configurations of composite particles belonging to the
27-plet of the flipped $E_6$ gauge group of symmetry; (c) represents a 
closed string describing a graviton.}
\lb{f3}
\efi

\clearpage\newpage
\bfi
\centering
\includegraphics[height=59mm,keepaspectratio=true]{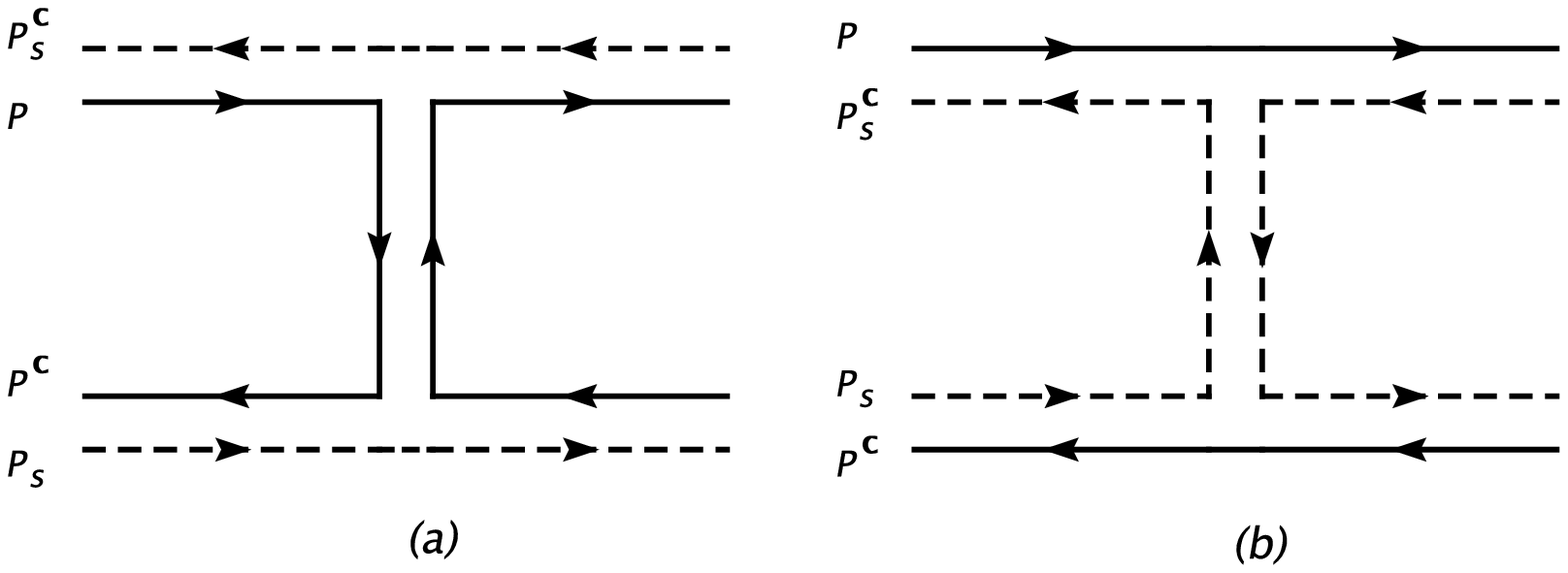}
\caption{Vector gauge bosons belonging to the 78 representation of the flipped $E_6$
and Higgs scalars -- singlet of $E_6$ -- are composite objects created 
(a) by fermionic preons $P,P^{\rm\bf c}$ and (b) by scalar preons $P_s,P^{\rm\bf c}_s$.
Both of them are confined by hyper-magnetic strings.}
\lb{f4}
\efi

\end{document}